\documentclass[acmsmall]{acmart}

\AtBeginDocument{%
  \providecommand\BibTeX{{%
    \normalfont B\kern-0.5em{\scshape i\kern-0.25em b}\kern-0.8em\TeX}}}

\usepackage[english]{babel}
\usepackage[utf8x]{inputenc}
\usepackage[T1]{fontenc}
\usepackage{amsmath}
\usepackage{graphicx}
\usepackage{booktabs}

\usepackage{hyperref}
\usepackage{xcolor}

\newcommand{\n}[1]{\textcolor{black}{#1}}

\setcopyright{acmlicensed}
\acmJournal{PACMHCI}
\acmYear{2021} \acmVolume{5} \acmNumber{CSCW2} \acmArticle{379} \acmMonth{10} \acmPrice{15.00}\acmDOI{10.1145/3479523}

\received{October 2020}
\received[revised]{April 2021}
\received[accepted]{July 2021}

\begin{document}

\title[Trend Alert: A Cross-Platform Organization Manipulated Twitter Trends]{Trend Alert: A Cross-Platform Organization Manipulated Twitter Trends in the Indian General Election}


\author{Maurice Jakesch}
\affiliation{
  \institution{Cornell Tech, Cornell University}
  \city{New York}
  \country{USA}
}
\email{mpj32@cornell.edu}

\author{Kiran Garimella}
\affiliation{%
  \institution{Massachusetts Institute of Technology}
    \city{Boston}
    \country{USA}
}

\author{Dean Eckles}
\affiliation{%
  \institution{Massachusetts Institute of Technology}
\city{Boston}
  \country{USA}
}

\author{Mor Naaman}
\affiliation{%
  \institution{Cornell Tech, Cornell University}
  \city{New York}
    \country{USA}
}

\renewcommand{\shortauthors}{Maurice Jakesch et al.}

\begin{abstract}
Political organizations worldwide keep innovating their use of social media technologies. 
In the 2019 Indian general election, organizers used a network of WhatsApp groups to manipulate Twitter trends through coordinated mass postings. We joined 600 WhatsApp groups that support the Bharatiya Janata Party, the right-wing party that won the general election, to investigate these campaigns. We found evidence of 75 hashtag manipulation campaigns in the form of mobilization messages with lists of pre-written tweets. Building on this evidence, we estimate the campaigns' size, describe their organization and determine whether they succeeded in creating controlled social media narratives. Our findings show that the campaigns produced hundreds of nationwide Twitter trends throughout the election. Centrally controlled but voluntary in participation, this hybrid configuration of technologies and organizational strategies shows how profoundly online tools transform campaign politics. Trend alerts complicate the debates over the legitimate use of digital tools for political participation and  may have provided a blueprint for participatory media manipulation by a party with popular support.
\end{abstract}





\maketitle


\section{Introduction}
In what has been called ``India's first WhatsApp election''~\cite{murgia_2019}, the world's largest democracy appointed a new parliament in 2019. With more than 400 million users, India has become WhatsApp's biggest market \cite{singh_2019} where social media has emerged as a key political battleground \cite{chakravartty2015mr, ahmed20162014, jaffrelot2015modi}. 
In preparation for the 2019 general election, the ruling Bharatiya Janata Party (BJP) and the contending Indian National Congress have invested massively in their social media operations \cite{jaffrelot2020bjp}. The parties built a country-wide network of social media organizers and claimed to have set up tens of thousands of WhatsApp groups to reach potential voters~\cite{goel_2018}. In the weeks before the election, reports emerged suggesting these groups were used to coordinate social media manipulation campaigns~\cite{dixit_2017}. However, prior accounts rely on indirect inferences~\cite{dixit_2017} or single instances of leaked evidence~\cite{russell_2019} as examining events that span multiple platforms is challenging.

The current study provides a first systematic analysis of the cross-platform campaigns. We document a novel configuration of technologies and organizational forms that shows how profoundly online tools are changing the nature of political organizing. We ask: How do \n{organizers} blend new and traditional technologies and strategies to shape the public agenda~\cite{karpf2012moveon}? What forms of political membership and participation emerge when traditional parties adapt to a \textit{hybrid media system} spanning multiple platforms~\cite{chadwick2017hybrid}? How effective are participatory media manipulation campaigns when supported by established organizations, and how do such campaigns differ from previously discussed instances of political astroturfing~\cite{zhang2013online, walker2014grassroots, king2017chinese, keller2017manipulate}?  

To begin exploring these questions, we joined more than 600 public WhatsApp groups that support the BJP. In these groups, we found evidence of~75 cross-platform media manipulation campaigns in the form of \textit{trend alert} messages. \n{These \textit{trend alerts}} messages call on group members to participate in \n{targeted} mass-postings on Twitter. The messages contain a link to a Google Doc where participants find a list of pre-written tweets to post at a specified time. We also acquired campaign-related Twitter data to analyze the campaigns. 
We find that that the campaigns were smaller than what media reports and claims by party officials may have suggested. Still, they produced nationwide Twitter trends through the voices of loosely affiliated online supporters. We estimate that hundreds of trends were produced throughout the election and show that other media outlets picked up the amplified narratives. While WhatsApp's relative anonymity makes it difficult to identify the campaign organizers, we provide evidence that the BJP's social media \n{department orchestrated} the campaigns in the background.

The campaigns complicate our understanding of social media manipulation and technology use by political organizations. 
\n{Their organizers} draw on the voices of a nationwide pool of online supporters and aggregate their micro-contributions through multiple platforms. \n{This} hybrid organizational strategy enables a low-cost ``permanent campaign''~\cite{neyazi2016campaigns} uniquely tailored to the capabilities of political parties with popular support. Centrally orchestrated but voluntary in participation, whether \textit{trend alerts} give rise to new forms of political propaganda or authentic participation is difficult to judge~\cite{jack2017lexicon, starbird2019disinformation}. Single-platform studies and policies will struggle to respond to such cross-platform manipulation campaigns. Furthermore, as election campaigns worldwide learn from each other, researchers need to watch how political organizations use media technologies outside the West. We argue that U.S.-centric  media policymaking may contribute to the weakening of political institutions \n{elsewhere}. 

\section{Background}
The current study expands on prior research on social media manipulation and technology use by political organizations. We summarize the relevant literature below and highlight our study's respective contributions. As many members of our community are not familiar with Indian politics, we provide a very short introduction to the Indian political context. 

\subsection{The Indian political context}
The Republic of India is the world's most populous democracy. From 11 April to 19 May 2019, India elected its 17th Lok Sabha, its bicameral parliament's lower house. The prime minister for the next five years was in turn chosen by the party with the most seats. The two main parties are the right-wing Bharatiya Janata Party, the BJP, and the left-wing Indian National Congress (INC). 

The need to reach young and increasingly connected voters ~\cite{singh_2019, tiwari_2019} has made social media a key political battleground in Indian politics \cite{chakravartty2015mr, ahmed20162014, jaffrelot2015modi}. India is  WhatsApp's largest market~\cite{singh_2019}, with over 400 million users on the platform. 
Surveys in India~\cite{lokniti2018} and Brazil~\cite{reuters2019report} have shown that about one in six users is part of political WhatsApp group reach a new demographic of voters.
In addition, thirty-four million Indians are on Twitter, with elite, urban, educated, and English-speaking demographics. With India being Twitter's fastest-growing user base~\cite{chaturvedi_2017}, this is changing rapidly. 
Since the 2014 Indian general election \cite{chakravartty2015mr, ahmed20162014, jaffrelot2015modi} all major political parties have invested heavily in their social media presence~\cite{tulasi2019catching, ahmed20162014, rajput2014social, goel_2018}, but media reports suggest they could not match the BJP's infrastructure~\cite{forbes2019socialmedia}.

The BJP claimed a landslide victory~\cite{safi_2019} and reappointed Narendra Modi as prime minister. The election results were seen as an affirmation of Narendra Modi's politics, whose political program blends arhetoric of managerial efficiency with the conservative ideology of Hindu hyper-nationalism \cite{khaitan2020killing}. Central to the ``NaMo'' brand -- an acronym for Narendra Modi -- is a focus on digital technologies for political outreach. Through significant investments into social media, Modi's team has created a positive image of a modern and approachable leader in the Indian public~\cite{pal2019making, chakravartty2015mr}. Modi accounts reach millions of followers on Twitter, LinkedIn, and Facebook, where he emerged as the second-most popular politician after Barack Obama \cite{pal2019making}. Mastering different platforms \cite{sinha2018fragile}, Modi's social media campaigns saturate the public space with favorable narratives \cite{jaffrelot2015modi} while bypassing the party traditional apparatus and critical press. Modi has also been accused of systematically undermining secularism, political pluralism, and intellectual freedom in India \cite{khaitan2020killing, goel_gettleman_khandelwal_2020, ganguly_9247}.  

\subsection{Technology use by political organizations}
Political organizations have traditionally occupied a central place in political participation. Organizations accumulated and spent resources, created collective identities, and pursued action programs~\cite{benford2000framing, jenkins1983resource, tarrow2011power}. According to ``Social Movement Theory 2.0'' \cite{shirky2008here} the emergence of the web changed this. The Internet reduces the cost of political organizing~\cite{margetts2015political} and provides a platform for large-scale social movements, such as the 2011 political uprising in Egypt~\cite{starbird2012will,wulf2013ground} or the Occupy Wall Street protests~\cite{tremayne2014anatomy}. Online political movements seem to rise from the grassroots with little or no formal organization~\cite{shirky2008here}, engaging millions of participants, while platforms take on the role of organizing agent~\cite{bennett2012logic}. 

Yet, even if online actions that succeed can reach millions, most political initiatives online fail \cite{margetts2015political}. Critics have long doubted that loose digital networks have the endurance and resourcefulness that is necessary to pursue long-term political change~\cite{tilly2004social, milan2015mobilizing}. In `The MoveOn Effect'~\cite{karpf2012moveon}, David Karpf argues that rather than \textit{organizing without organizations}, the new media environment has given rise to \textit{organizing through different organizations}. The Interned altered one set of organizing constraints, while another set of political fundamentals remained unchanged \cite[p.~8]{karpf2012moveon}. Rather than replacing the organization, the shrinking costs of political communication have transformed the \textit{organizational layer} of politics, producing new tactical repertoires and new workflows within political organizations. 

Political campaigns like MoveOn.org experiment with new forms of organizational structure, membership, engagement practices, and fundraising. At the same time, they rely on a central organization to formulate an agenda and coordinate efforts ~\cite{karpf2012moveon}. The varying requirements of different media systems give rise to what Andrew Chadwick calls \textit{hybrid organizations} \cite{chadwick2017hybrid}. In a \textit{hybrid media system}, older and newer technologies, tactics, and organizational forms co-evolve based on adaptation and interdependence \cite[p. 25]{chadwick2017hybrid}. \textit{Hybrid organizations} succeed in rapidly changing environments by combining multiple, loosely coupled groups and temporal instances of action \cite[p. 28]{chadwick2017hybrid}. 

Our analysis traces a hybrid configuration of technologies and organizational forms that turns online supporters into a manageable political resource.

\subsection{Social media manipulations during elections}
The tools and scale of social media have opened new doors to media manipulation \cite{bradshaw2019global, woolley2018computational, lazer2018science, allcott2019trends} and have raised concerns about election influence worldwide \cite{grinberg2019fake, bessi2016social, howard2016bots, woolley2018computational}. Social media manipulations in India, however, have received \n{comparatively} little academic attention. Anupam and Schroeder \cite{das2020online} have shown that Indians were concerned about disinformation campaigns and mediated religious extremism during the 2019 general election. \citet{udupa2015internet} analyzed how right-wing Hindu nationalists make tactical use of online tools, while \citet{chaturvedi2016troll} demonstrate how hate speech by right-wing trolls incited online communal tensions. Disinformation campaigns in India also have a distinct cultural component as tactics often blend modern narratives with Hindu mythologies~\cite{rajan2019new}.

The current study draws on prior research on political astroturfing. Astroturfing is a genre of social media manipulations where organizations seek to influence debates by creating the impression of a genuine citizen movement~\cite{zhang2013online, walker2014grassroots}. The 50 Cent Army~\cite{king2017chinese,yangPennyYourThoughts2015}, a  group of social media users paid by the Chinese government, is a well-known example of astroturfing. Studies have also described astroturfing campaigns by the South Korean National Information Service \cite{keller2017manipulate} and by users identified as Russian and Iranian state-sponsored trolls \cite{zannettou2019let}. Early studies on astroturfing assume that campaigns rely on automation and inauthentic user accounts \cite{ratkiewicz2011detecting}, but more recent work argues that astroturfing research needs to consider human involvement more carefully \cite{keller2020political, ferrara2016rise}.

Information manipulation studies tend to focus on behaviors on a single media platform, such as Twitter. Only recently researchers have started to report on information operations spanning multiple platforms, where organizers mobilize users on one platform to influence debates or threaten individuals on other media \cite{wilson2020cross, bradshaw2017troops}. For example, \citet{zannettou2018origins} showed how groups from fringe websites like 4chan and Gab influence agendas on other social media platforms during the 2016 U.S. elections.  \citet{wilson2020cross} have explored how Twitter and YouTube are used in complementary ways to create doubt about the Syrian regime's atrocities in the Syrian civil war. While some cross-platform campaigns are hostile takeovers of a debate through a group from another platform, others deliberately combine the affordances of different platforms to reach their goals~\cite{phadke2020many}.

There is little prior empirical research on large-scale cross-platform manipulation. Cross-platform campaigns are \n{difficult} to study, as their analysis requires data collection on multiple platforms and a link between activities on different platforms. The current study documents an extensive cross-platform media manipulation campaign in India that resembles but differs from astroturfing.

\section{Methods}

\begin{figure}
        \includegraphics[width=0.75\textwidth]{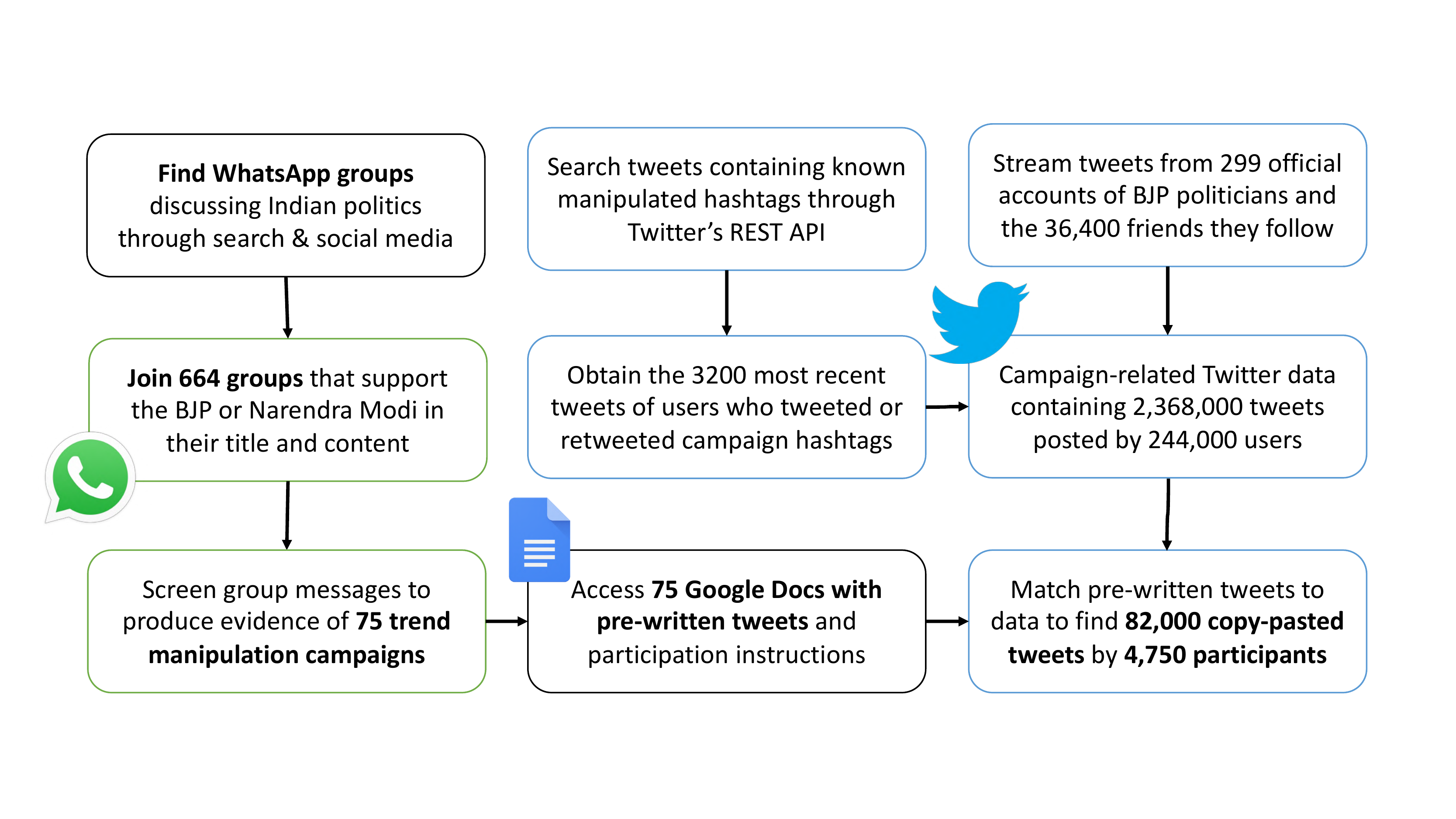}
      \caption{Overview of the data collection procedure across WhatsApp, Google Docs and Twitter}
  \label{fig:collection}
\end{figure}

Our data collection \n{spans} multiple platforms to provide a broad, multifaceted understanding of how the campaigns are organized, who participates in them, and whether they are effective. Figure~\ref{fig:collection} shows a summary of the data collection process. 

To start investigating, we looked for WhatsApp groups in India that discuss politics. 
We searched for links containing ``\url{chat.whatsapp.com}'' along with a list of keywords related to Indian parties and their political leaders. Lists of such public groups are advertised through social media and dedicated websites, for example, \url{https://whatsgrouplink.com/}. We manually screened the results to remove WhatsApp groups unrelated to Indian political news. To constrain the scope of relevant data, we focussed the analysis on the groups of one party only. We chose to analyze WhatsApp groups that supported the BJP as the BJP dominates Indian social media \cite{pal2019india} and won the 2019 general election. Additional analyses of other parties' WhatsApp groups will produce a more comprehensive picture. 

We joined 664 WhatsApp groups whose titles and content supported the BJP or Narendra Modi using tools provided in prior work~\cite{garimella2018whatsapp}. The groups are a convenience sample and may not be representative of the BJP's groups.  Members identify themselves through phone numbers only, so we cannot determine whether a group is administrated by party staff or by independent supporters. Since the groups can be accessed through publicly available links, there is no reasonable expectation of privacy. We declared ourselves as researchers and explained that the account was used for research purposes in our WhatsApp profiles. 

We searched the groups for evidence of the social media campaigns reported on by journalists \cite{dixit_2017, russell_2019}. Specifically, we looked for messages that referred to a Twitter hashtag and linked to a Google Doc. We found evidence of 75 manipulated Twitter hashtags.
To analyze the 75 campaigns, we collected tweets containing the campaign hashtags. We used multiple data sources to achieve high coverage of campaign-related activities on Twitter: 
\n{Building on prior work by \citet{tulasi2019catching},
we identified 299 official party accounts and accounts of elected representatives. The list of these accounts is included in the appendix. From January 2019 onwards, we collected their tweets through Twitter's streaming API. We also collected tweets posted by the accounts the politicians followed (150k friends in total, of which 36,446 were unique). }

In August 2019, we queried the Twitter search API to find tweets with campaign hashtags. For each retrieved tweet, we also collected the 100 most recent retweets\footnote{The hard limits are imposed by the Twitter API.}. Finally, we obtained the 3,200 most recent tweets of the official BJP accounts, the accounts they followed, and all users who had tweeted or retweeted the campaign hashtags. The campaign-related Twitter data set consists of 2,368,000 tweets posted by 244,000 users.

We purchased historical trend data from the Twitter Trending Topics Archive.\footnote{\url{https://rapidapi.com/onurmatik/api/twitter-trending-topics-archive}} The data set contains snapshots of Twitter's list of trending topics taken in half-hour intervals in cities across India. It also contains tweet counts for the trending hashtags provided by Twitter, which we compared to our data to estimate our coverage. We estimate that we obtained 50\% to 80\% of all available tweets and retweets for manipulated hashtags. We assume that the missing tweets are largely retweets that we could not access due to Twitter's austere limits on retweet retrieval. As our analysis is \n{focused on} original tweets, missing retweets does not affect our findings. We may have also missed older tweets by accounts posting more than 3,200 tweets and tweets by deleted accounts.

We linked the Twitter data to the pre-written tweets found in Google Docs. Although campaign organizers asked participants to modify the pre-written tweets, many posted them as they appeared in the Google Doc. If they made changes, modifications were typically minor, such as adding hashtags or user mentions, changing punctuation, or replacing a single word.  Before matching the collected tweets to the pre-written tweets, we removed hashtags, URLs, user mentions, and non-word characters. In addition to exact matching, we performed space-less matching, as some spaces were lost when pre-written tweets were copied. We also calculated a fuzzy match to identify tweets where up to five characters have been modified. We only considered tweets longer than 20 characters for exact matching or 50 characters for fuzzy matching. We found 82,200 original tweets with pre-written content.

The MIT Institutional Review Board has approved the data collection and analysis on WhatsApp. The Twitter data collection and analysis have been approved by the Institutional Review Board at Cornell University. To protect users' privacy, we do not publicize usernames for non-official accounts with less than 100,000 followers. 

\section{Results}
To illustrate the interplay of the technologies and organizational forms in the campaigns, we first provide a case study of the ``Modi Mein Hai Dum'' hashtag. Next, we describe the cross-platform campaign infrastructures in detail to show how the organizers blend novel and traditional technologies and organizational strategies to shape the public agenda. Finally, we analyze whether the participatory media manipulations succeeded in producing Twitter trends. 

\subsection{Case study: The ``Modi Mein Hai Dum'' campaign} 
\begin{figure}
  \begin{center}
    \includegraphics[width=0.37\textwidth, trim=0 60 0 00,clip]{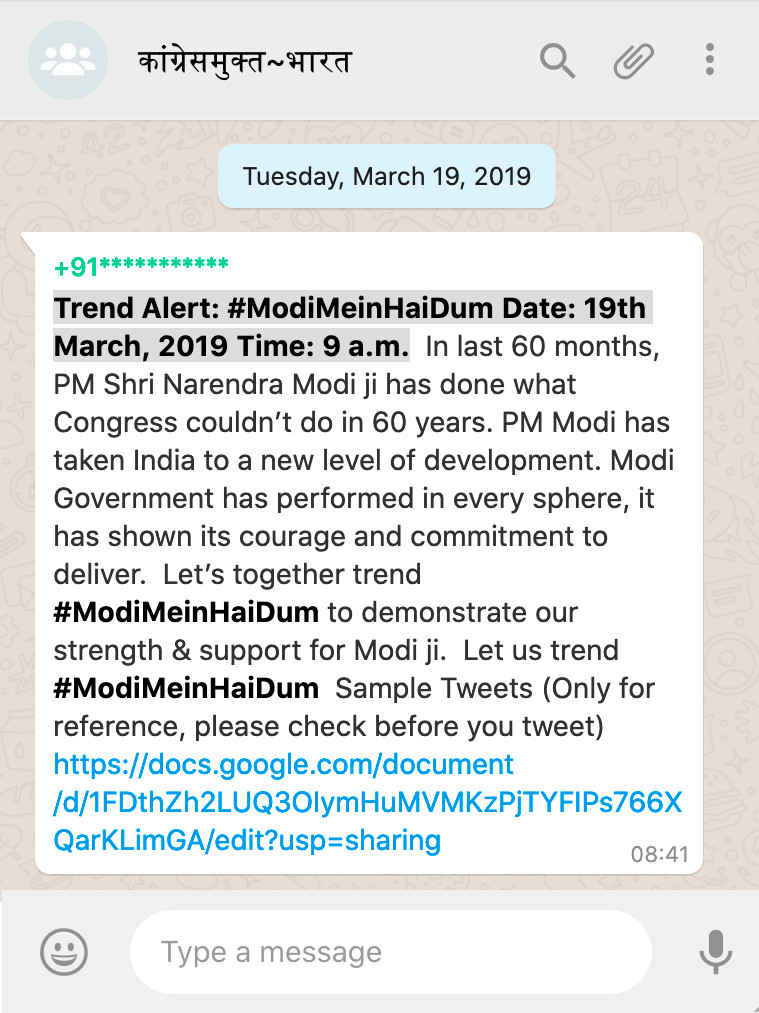}
  \end{center}
\caption{A \textit{trend alert} mobilizes WhatsApp group members for the \#ModiMeinHaiDum campaign}
\label{fig:modimeinhaidum}
\end{figure}

Three weeks before the general election, a message was posted to a WhatsApp group we joined: ``Trend Alert: \#ModiMeinHaiDum Date: 19th March, 2019 Time: 9 a.m.''.
The hashtag \#ModiMeinHaiDum means that Narendra Modi, the BJP's iconic leader figure, is strong and committed. Figure~\ref{fig:modimeinhaidum} shows a screenshot of the message.
After explaining the context of the campaign, the organizer called the  250 members of the group to action: ``Let's together trend \#ModiMeinHaiDum to demonstrate our strength \& support for Modi Ji. Let us trend \#ModiMeinHaiDum!''  In this context \textit{trending} refers to synchronously posting a large volume of tweets with a specified hashtag to create a Twitter trend.

\begin{figure}            
\includegraphics[width=0.85\textwidth]{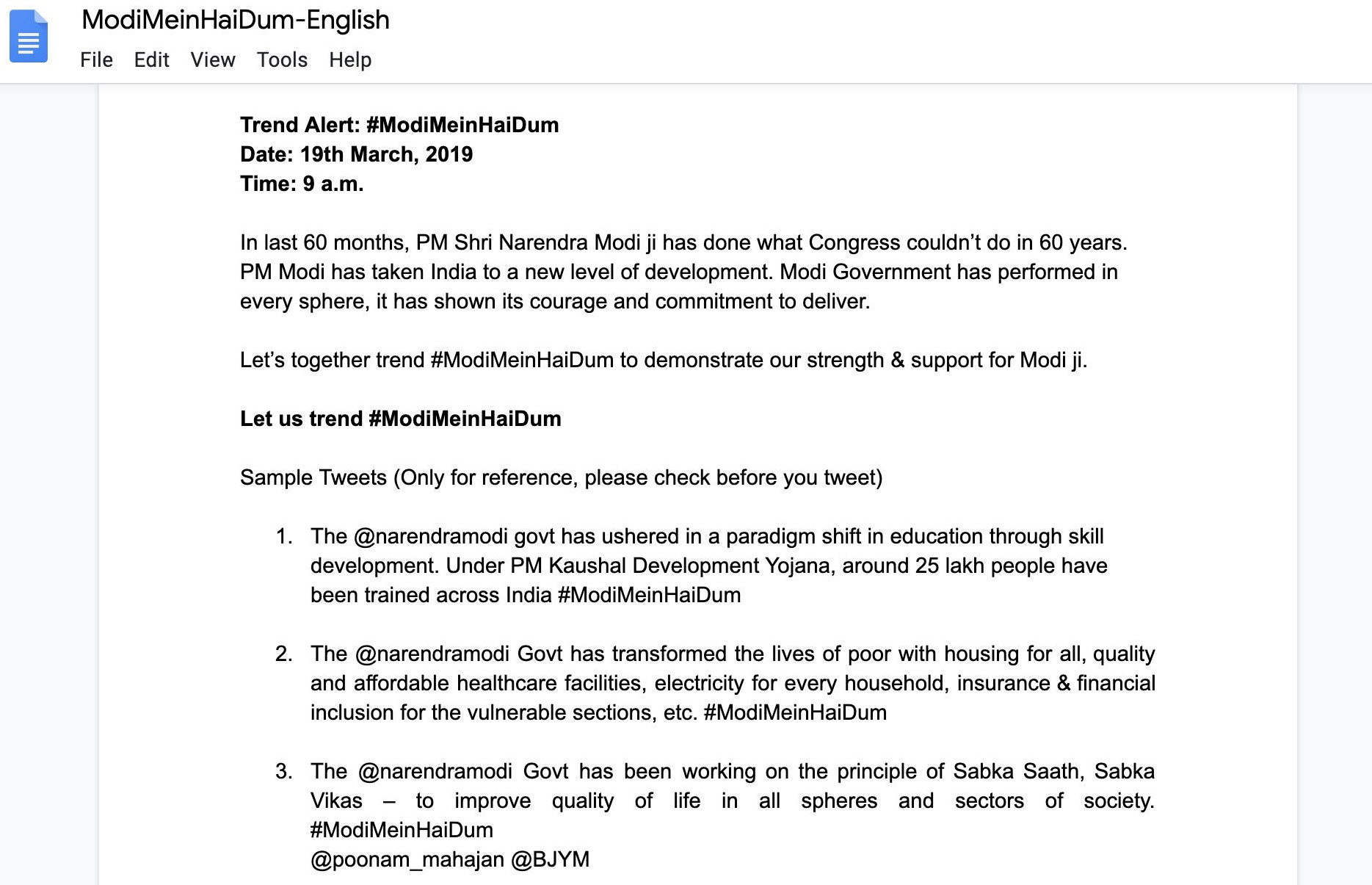}
      \caption{A \textit{tweet bank} hosted on Google Docs frames the campaign narrative with pre-written tweets}
  \label{fig:document}
\end{figure}

\begin{figure}
        \includegraphics[width=0.9\textwidth]{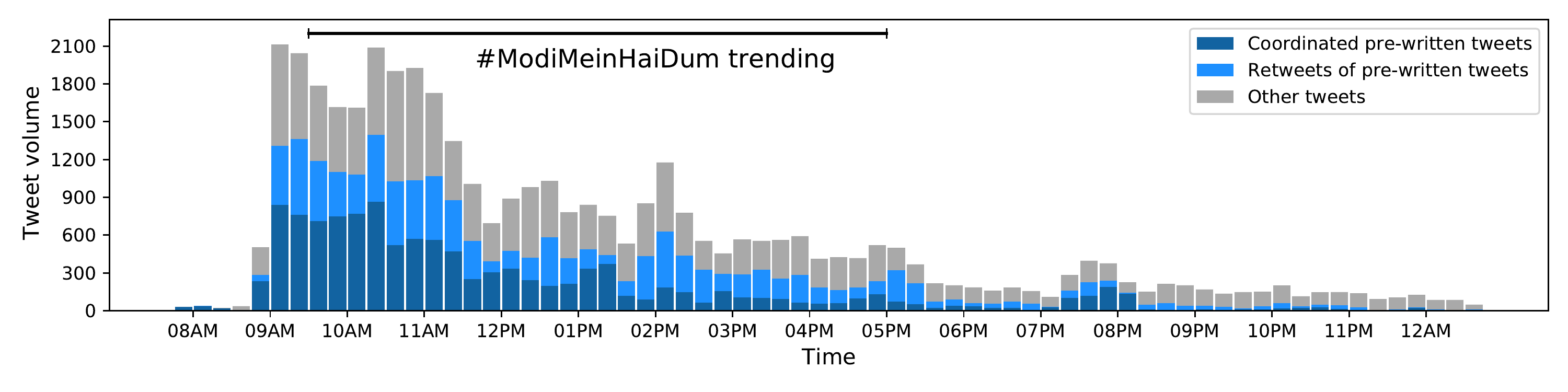}
        \caption{The campaign started with 2,100 tweets per 15-minute interval at 9:00AM. }
        \label{fig:case-volume}
\end{figure}

\begin{figure}
        \includegraphics[width=\textwidth, trim={0 2.5cm 0 1.5cm}, clip]{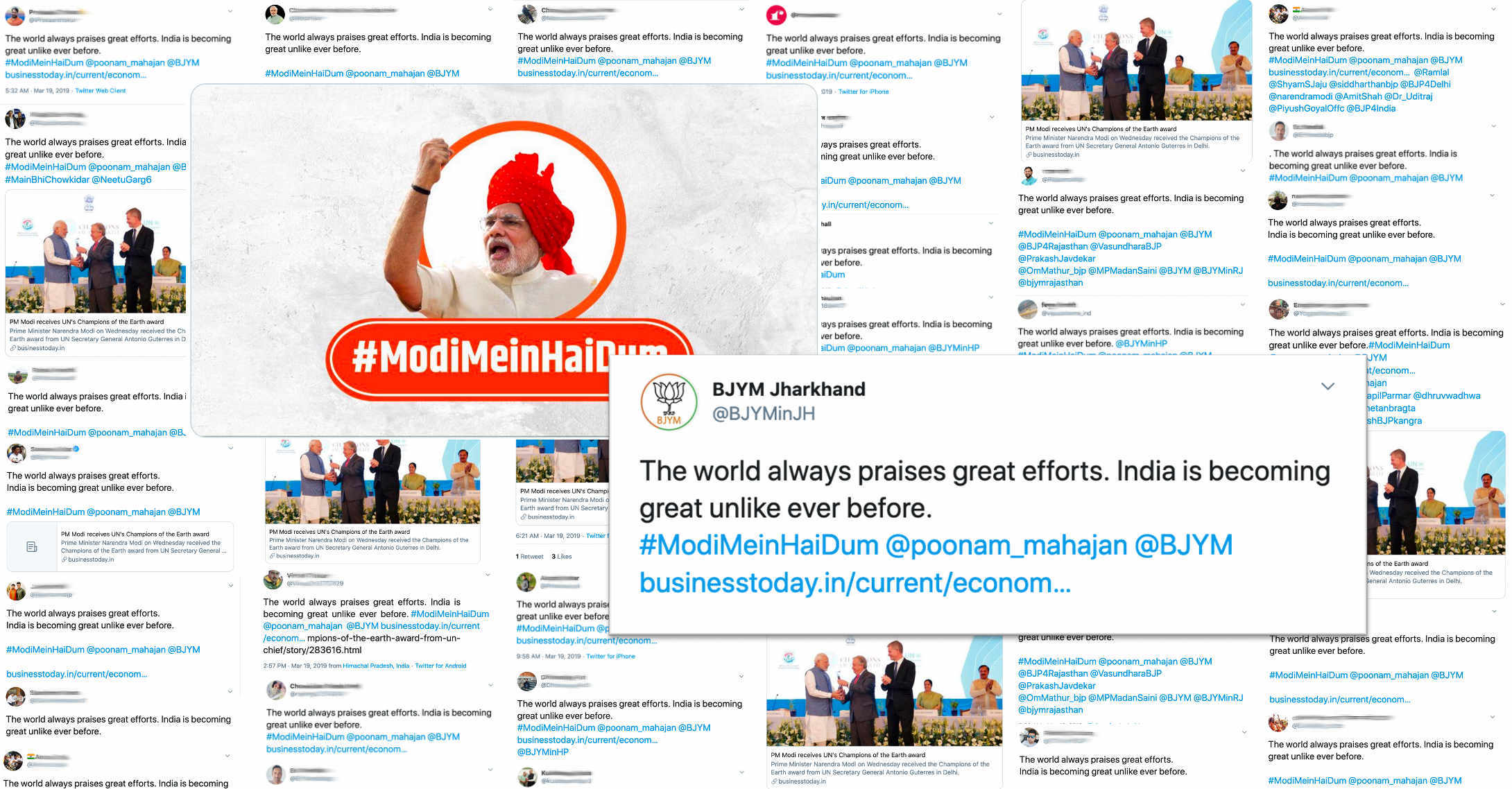}
      \caption{Tweets from the $\#$ModiMeinHaiDum campaign with identical content.}
  \label{fig:case}
\end{figure}
The organizer also linked a so-called \textit{tweet bank}, a document hosted on Google Docs where supporters find further participation instructions and a list of 100 pre-written tweets (see Figure~\ref{fig:document}). The \textit{tweet bank} makes it easy to participate in the mass posting and allows the organizers to control the campaign's narrative. 

The first \#ModiMeinHaiDum tweet was posted at 7:47 a.m. Indian Standard Time. It matches the first pre-written tweet in the GoogleDoc. The posting user identified himself as a Himachal Pradesh IT cell member, a regional chapter of the BJP's social media arm. 
Around 8:50 a.m., minutes before the designated launch time, the campaign took off. By 9:05 a.m., 68 users had posted almost 500 template tweets. Figure~\ref{fig:case-volume} illustrates the tweet volume during the  \#ModiMeinHaiDum campaign. At 9:30 a.m., the campaign has reached a volume of 4,800 tweets and \#ModiMeinHaiDum trends across India. \#ModiMeinHaiDum remained on  Twitter's trend list for eight hours until 5~p.m. By the end of the day, the campaign had accumulated 46,000 tweets, and 416 users had posted tweets from the \textit{tweet bank}. Figure \ref{fig:case} shows a sample of these tweets. 

\subsection{Cross-platform campaign infrastructures}
Building on the initial view of a single campaign provided in the case study, we now expand our analysis to a larger number of campaigns. Through multiple lenses, we explore how the organizers created a hybrid cross-platform organization that turned loosely affiliated online supporters into a political resource. 

\subsubsection{Support repositories: WhatsApp groups and the NaMo app} 

Our sample of WhatsApp groups is neither exhaustive nor representative, as their off-the-record nature makes them challenging to study. While incomplete, the characterization it provides a first impression of the support repositories that the organizers draw on for their campaigns.

The WhatsApp groups \n{where the campaigns were advertised} have names and political content related to supporting the BJP. Some \n{groups} offer news, like ``News group of BJP'' or ``Namo Broadcast -3''. Some explicitly gather party supporters, like ``BJP social support'', ``BJP Defense Team Tvlr Wst'', or the ``BJP Cyber Army 350+''. Others, like ``BJP BELLARY'' or ``BJP GULBARGA'' represent regional party chapters. We cannot ascertain the total number of groups, but we identified a cluster indexed based on location. Groups for smaller towns rise from ``002 FTNBJP Ponneri'' to ``226 FTNBJP Palayamkottai'' and larger cities have their dedicated indexing, such as ``FTNBJP 05 Chennai South''. This naming convention would suggest that the state of Tamil Nadu may have far fewer than the 50,000 groups claimed by journalistic reports and party officials~\cite{goel_2018}.

\begin{figure}
        \includegraphics[width=0.9\textwidth, trim={1cm 0 1cm 0}]{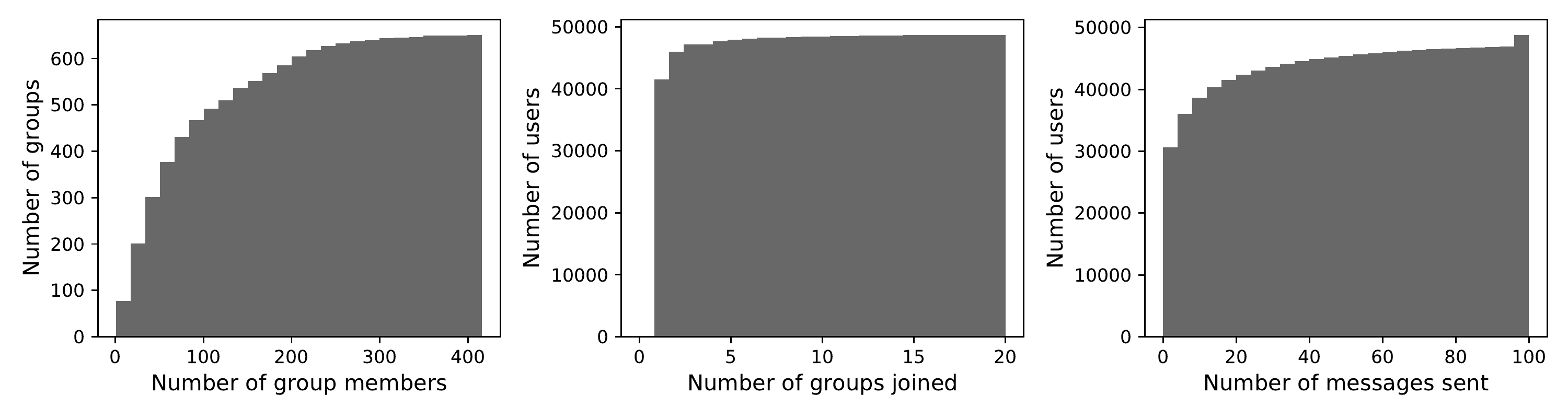}
        \caption{Most group members joined only a single group and do not engage in conversations.}
        \label{fig:groups}
\end{figure}

WhatsApp groups present an organizational innovation in the membership structure of political organizations. While some users have joined more than 200 groups, most users are members of only one group. These members may not have an official party affiliation and may be unaware that they have become part of an organization. Still, they make up a repository of almost 50,000 potential supporters that can be called upon for political action. Figure \ref{fig:groups} shows member statistics. 

Three admins run the median group. WhatsApp requires no identification beyond a phone number, allowing the organizers to broadcast through an opaque channel while remaining largely anonymous. Some admins posted messages to different groups within milliseconds, hinting at automation tools and a coordinated organizational effort behind seemingly separate groups. We speculate that many admins and content creators are part of the BJP IT cell~\cite{chaturvedi2016troll}, the social media arm of the party.  In addition to a core team in Delhi, the IT cell draws on a large network of regional chapters across the country~\cite{udupa2019nationalism} \n{that could administrate a federated network of WhatsApp groups.} 

The most evident link between the campaigns and the BJP is the party's official Narendra Modi app. The app features ``messages and emails directly from Prime Minister Narendra Modi'' and opportunities to ``contribute towards various tasks''. These volunteer tasks include participation in hashtag manipulation campaigns we observed. According to Google Play, the app has been installed more than 10 million times.\footnote{\url{https://perma.cc/TZ2K-4Q8V}}

\subsubsection{Mobilizing group members through \textit{trend alerts}} 
In the groups we joined, we found \textit{trend alerts} for 75 campaigns, broadcast to the group like an advertisement. They start with the campaign hashtag, the date, and the time of the action, e.g., ``\#ForTheFirstTime  Time: 9.00 a.m. Date: 8th February 2019''. \textit{Trend alerts} then provide context for the campaign, such as a politician's visit or the failings of the rival Congress party:
\begin{itemize}
\item[>>] \textit{``Let us ensure that the nation continues to progress under Modi ji. Let's make sure that Modi Ji comes back to power again in 2019.  One nation. One vice. One aim. Modi ji again!  Let's trend to mobilize youth from all across the nation!''}
\item[>>] \textit{``Rahul, Kejriwal [rival party politicians] and other opposition parties lied on Rafale [a controversial deal with a French airplane manufacturer] and insulted the country on a global platform. Let's expose their lies. Everyone is requested to please support the trend \#RaFailGandhi.''}
\item[>>] \textit{``This [BJP government program] is not spoken about much in the media. Hence today we are trying to bring about awareness regarding the immense work done in MODIfying cities throughout India using the hashtag \#MODIfiedCities!''}
\end{itemize}

\textit{Trend alerts} also link to a \textit{tweet bank} or \textit{image bank}:
\begin{itemize}
\item[>>] \textit{``For sample tweets reference : \url{https://docs.google.com/document/***} Note - Please don't just copy paste the sample tweets, please alter it a bit.''}
\item[>>] \textit{``There are 27 graphics, 4 illustrations, 2 collages, 1 NIJ video and tweets which can be spread. This material is also available in readily spreadable format in Your Voice section in the Volunteer Module of the Narendra Modi App.''}
\end{itemize}

We found no evidence of rewards or payments for participants, indicating that campaign participation was voluntary. Instead, some \textit{trend alerts} thank participants or create pressure to participate: \textit{``We highly appreciate your kind support and co-operation. Thank You!!''} or \textit{``Your cooperation is expected.''}

Most \textit{trend alerts} come from users sending only a single call to action. We interpret their posts as forwarding from other channels rather than a central entity's systematic mobilization.  However, we also identified a single phone number that posted 42 \textit{trend alerts} with timings hinting at the use of automation tools and a more centralized effort. 


\subsubsection{Setting the campaign narrative through \textit{tweet banks}} 

\textit{Tweets banks} contain a mean of sixty pre-written tweets that facilitate participation and allow the organizers to steer the campaign narrative. Many emphasize the achievements of the ``Modi Government'' and praise Narendra Modi's personal qualities. Others accuse the rivaling Congress party of various failures and express adoration of India, its army, and products. 
We found no evidence of political agitation or hate speech against minority groups in the \textit{tweet banks}. We did not fact-check the pre-written tweets but found that more than half of the templates use numbers and statistics to make a claim.
The sample of pre-written tweets below gives an impression of their style and content: 

\begin{itemize}
   \item[>>] \textit{Modi Govt. declared Polavaram project as project of ``National Importance'' -Modi Govt. Funding 100\% cost of Polavaram Project -About 7000 Cr already given by Modi Govt. for Polavaram Project \#BJP4BetterAndhra}
    \item[>>] \textit{\#MizoramWithModi Congress brought ``Vote for Money'' culture in Mizoram, infiltrating the minds of the poor tribals in Mizoram.}
    \item[>>] \textit{PM @narendramodi will initiate distribution of looms and frames to carpet weavers. 10,000 weavers will receive them in next 2 months. \#MODIfiedTextiles}
    \item[>>] \textit{The verdict of the Supreme Court is a slap on all such attempts to mislead the country. \#RaFailGandhi}
    \item[>>] \textit{We all grew up in awe of the army and their bravery. It makes sense for us to celebrate our soldiers on \#ArmedForcesFlagDay. Let us honour their bravery by contributing to the Armed Forces Flag Day Fund and wearing the armed forces flag.}
\end{itemize}

For 42 of the 75 investigated campaigns, the tweets we found in the \textit{tweet banks} could also be posted through the volunteer section of the official Narendra Modi app. This connection proves the BJP's involvement in the campaigns and shows the extent to which hybrid manipulation campaigns cross the boundaries of platforms and organizations. 

\subsection{Campaign participation: Who contributed?}

\begin{figure}
        \includegraphics[width=0.9\textwidth, trim={1cm 0 1cm 0}]{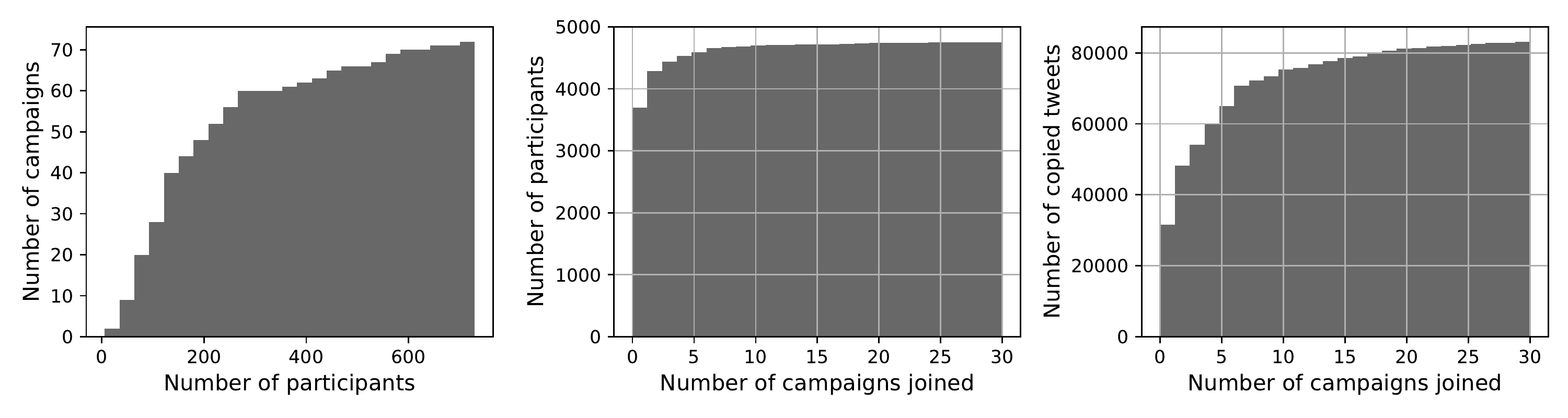}
        \caption{Most participants are loosely affiliated supporters joining only one or two campaigns.}
        \label{fig:contributions}
\end{figure}

Participation in astroturfing campaigns is difficult to ascertain, but we were able to identify Twitter users who participated in the campaigns. 
We classify users that posted original tweets (excl. retweets) that are identical to templates in the \textit{tweet bank} as campaign participants. In the average campaign, 141 participants copy-pasted tweets from the \textit{tweet bank}. 
Compared to the tens of thousands of WhatsApp groups claimed by media reports and party officials \cite{goel_2018} and considering that supporters were also engaged through the official party app, 141 participants is a modest campaign size. However, it is possible that some supporters participated without copying pre-written tweets.

Rather than size, what makes these campaigns innovative is the type of supporters they enroll. Occasional participants make up for the vast majority of participants and they contributed most posted tweets. Figure~\ref{fig:contributions} shows the distribution of contributions. Most supporters joined for only one or two hashtag manipulation campaigns, suggesting loose support driven by occasion and convenience rather than systematic involvement. Paid supporters and bot accounts would likely participate in more than one or two campaigns. Instead, trend alert campaigns are the work of a diverse and disparate group of online supporters. 

However, the campaigns also have a base of core supporters. The three percent of participants who participated in more than five campaigns posted 21\% of the pre-written tweets. We expect that these participants are more formally involved with the organization behind the campaigns. To get a sense of who they are, we manually screened profiles of the 50 users that participated in more than ten campaigns: 16 describe themselves as members of the BJP IT cell, the party's social media wing. Six accounts are official Twitter accounts of high-profile BJP politicians, such as members of parliament or state ministers. Eighteen more mention an affiliation with the BJP or the RSS, a right-wing Hindu nationalist volunteer organization. The remaining core supporters describe themselves as ``staunch NaMo supporter[s]'' and committed patriots but do not disclose an official party affiliation. Seven core supporters had their accounts suspended by Twitter.

The participation structures show that the campaigns and combine larger numbers of loosely affiliated supporters with the power of a more professional core of regular supporters.

\subsection{Evaluation: How successful were the campaigns?}
The campaigns were highly effective at producing lasting Twitter trends with a relatively small number of participants. Out of the 75 campaigns, 69 succeeded in reaching India-wide trend status. While Twitter's criteria for trend status seem to depend on various factors, most campaigns made it onto the trend list after accumulating about 5,000 tweets within 30 minutes. Not all of these needed to be copy-paste tweets from the \textit{tweet bank} -- some were retweets, or organic conversations spun off from coordinated activities. 

Campaigns that achieved trend status remained on Twitter's trend list for about ten hours. By the end of the trend, manipulated hashtags had accumulated a mean of 42,400 tweets. Twitter does not provide exposure data, but we found that some manipulated hashtags spread far beyond the circles of users who posted the templates. 32 of the 75 campaigns were picked up by other media outlets, such as Quartz India\footnote{\url{https://perma.cc/RS45-GSME}},  Financial Express\footnote{\url{https://perma.cc/EY76-LKCL}},  HuffPost India\footnote{\url{https://perma.cc/SH56-LN5J}},  The Hindu\footnote{\url{https://perma.cc/XMB7-ACUC}} and Firstpost\footnote{\url{https://perma.cc/67HV-XTTR}}.

\begin{figure}
        \includegraphics[width=0.68\textwidth]{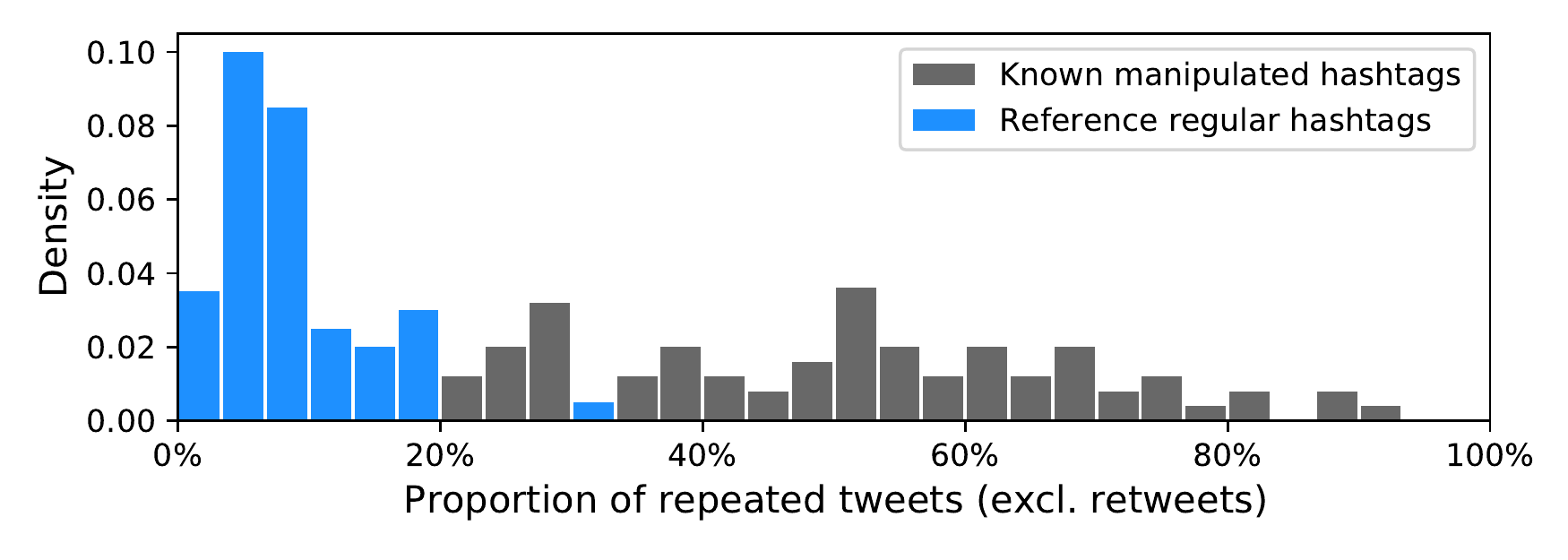}
        \caption{Content repetitions are highly characteristic of manipulated hashtags.}
        \label{fig:percentageduplicate}
\end{figure}

\begin{figure}
        \includegraphics[width=0.83\textwidth]{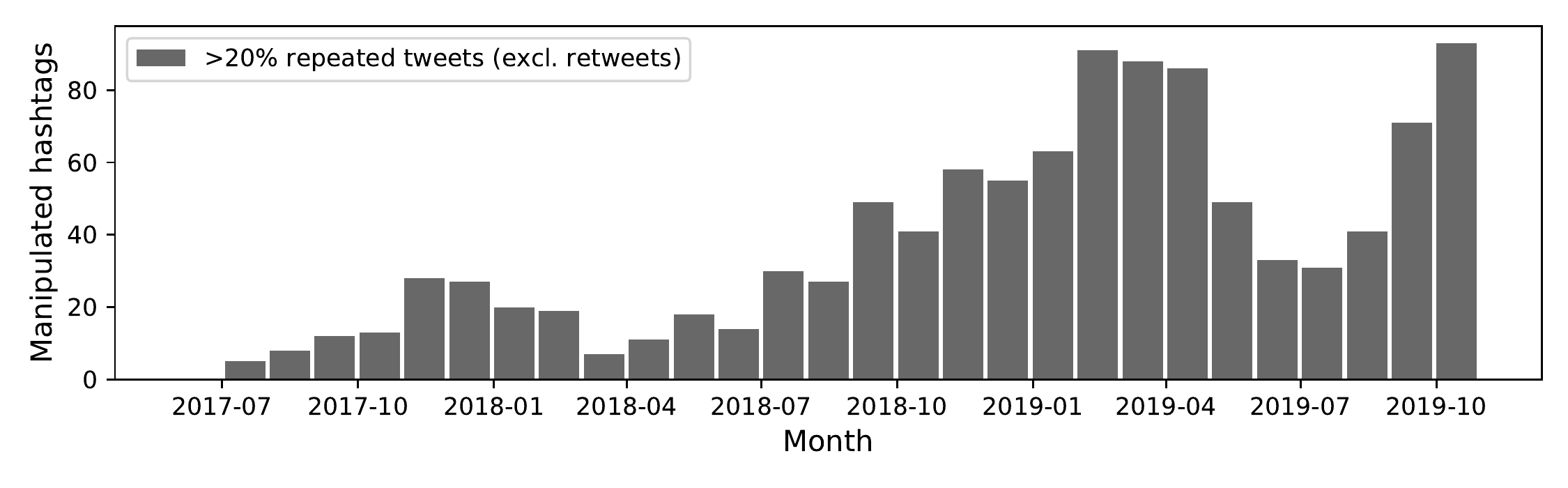}
        \caption{The estimated number of manipulated hashtags rises and falls with election dates.}
        \label{fig:totalscope}
\end{figure}

We finally estimated the total scope of the hashtag manipulations based on the observation that, for regular Twitter posts, the likelihood of tweets being identical is relatively low. When posting from a \textit{tweet bank}, however, the share of repeated tweets increases dramatically. To test this assumption, we collected a reference set of popular hashtags. We chose hashtags associated with the 50 largest cities in India (e.g., \#mumbai or \#delhi) as hastags that were unlikely to be manipulated. Figure~\ref{fig:percentageduplicate} shows the proportion of repeated content in city hashtags compared to known manipulated hashtags. Excluding retweets, almost none of the city hashtags contained more than 20\% repeated tweets. In comparison, all 75 manipulated campaigns we investigated had more than 20\% repeated content.

To estimate the total number of manipulated hashtags, we examined all hashtags with a volume of at least 500 tweets that were used by at least five users who participated in the 75 known manipulation campaigns. We find 1,131 hashtags that contain more than 20\% repeated content, which we take to as a strong indication that the posting was coordinated. With a more conservative threshold of at least 35\% of repeated content, we find 624 hashtags that were most likely manipulated.  

Figure~\ref{fig:totalscope} shows the number of estimated monthly campaigns.  The number of manipulated hashtags rises in November and December 2017, when state elections were held in Himachal Pradesh and Gujarat. It climaxes during the 2019 general election in February, March, and April. The campaign volume drops after the general elections but then rises again in fall 2019 during the assembly elections in Maharashtra and Haryana. The development shows that the \textit{trend alert} strategy can respond to events in real-time and organizers make extensive use of this capability  beyond the 2019 general election.

\section{Discussion}
We have described how organizers combined new technologies and novel forms of organization to run strategic information operations in the 2019 Indian general election. 
Our results provide an updated understanding of how profoundly online tools and platforms transform the \textit{organizational layer} of campaign politics.
The findings raise the critical question of whether political participation enabled by such hybrid campaigns is authentic and legitimate.
We argue that researchers need to study how political organizations use new technologies across countries and platforms to develop better theories of and responses to hybrid information campaigns.

\subsection{A blueprint for organizationally brokered media manipulation?}
\textit{Trend alerts} are participatory social media manipulation campaigns running on top of an organizational backbone \n{that combines} multiple platforms. While the WhatsApp groups seem informal, an organization (most likely the BJP's social media department) develops the campaign narratives and maintains the infrastructure in the background. This organizational backbone creates coherence and persistence of action that loose multi-issue networks can rarely generate~\cite{bennett2011digital}, differentiating the campaigns from the activities of self-organizing advocacy groups~\cite{shirky2008here} or trolling collectives~\cite{hine2017kek}. 
However, the campaigns also differ from orchestrated information operations such as the 50~Cent Army~\cite{yangPennyYourThoughts2015, king2017chinese} or social bots~\cite{bessi2016social}. 
 \textit{Trend alerts} are the collaborative work of multiple, loosely coupled groups and temporal instances of action. Participation in \textit{trend alerts} is largely voluntary. Instead of direct control, organizers monitor and nurture the campaigns from a close distance~\cite{udupa2019nationalism}, similar to public policy advocacy organizations like MoveOn.org. In such ``organizationally brokered networks of collective action''~\cite{karpf2012moveon}, a central entity ``mobilizes and manages participation and coordinates goals''~\cite{bennett2012logic} while loosely affiliated supporters carry out the actions.
 
 \begin{figure}
		\includegraphics[width=1\textwidth, trim={1cm 3.5cm 1cm 1.5cm}]{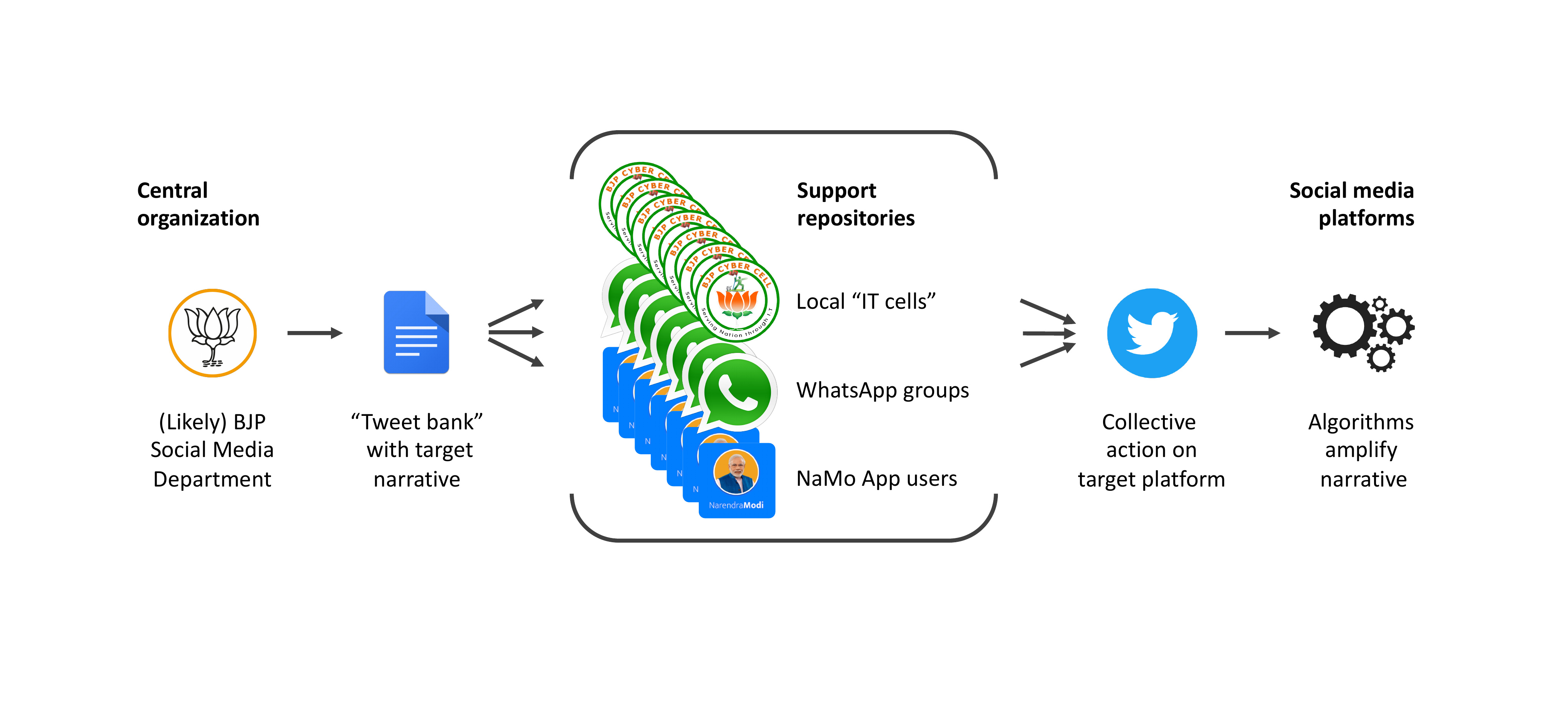}
	  \caption{The functional layout of the cross-platform organization for trend manipulation}
  \label{fig:process}
\end{figure}

As illustrated in Figure \ref{fig:process}, \textit{trend alert} campaigns exhibit a mastery of different platforms and hybridity of organization \cite{chadwick2017hybrid} that goes beyond what previous studies have observed. Where MoveOn.org runs an e-mail list to reach potential supporters, \textit{trend alert} organizers draw from a nationwide pool of WhatsApp support groups,  an official party app with millions of users, and a federated network of regional IT cells~\cite{udupa2019nationalism, chaturvedi2016troll} for their actions. 
Their usage of WhatsApp groups, in particular, is innovative, as the groups allow direct communication and interactions with supporters in micro-organizations while enabling the scale of an e-mail list at almost no additional cost. 
As opposed to advocacy organizations that are optimized to minimize running costs~\cite{karpf2012moveon}, \textit{trend alert} campaigns are most likely backed by one of the world's largest political parties with vast resources and traditional organizational presences across the country. 

When an established political party transforms itself into an organizational hybrid to use the full potential of online organizing, it can run information operations of a scale and pervasiveness that surpasses what centrally controlled campaigns~\cite{yangPennyYourThoughts2015, king2017chinese} and self-organizing advocacy groups~\cite{shirky2008here} achieve. \textit{Trend alerts} allow an organization to continuously reenact, promote and defend its narratives on social media, enabling a ``permanent campaign''~\cite{neyazi2016campaigns} at low cost. In our data, we anecdotally observed other Indian parties trying to replicate the method. We believe that \textit{trend alerts} may have provided a blueprint for participatory media manipulation by political parties with popular support.

\subsection{Astroturfing with genuine support: Legitimacy and authenticity}
We need to consider whether the described campaigns are legitimate, acceptable to the public, and compatible with platform policies. At the surface level, the campaigns resemble the astroturfing campaigns described elsewhere~\cite{king2017chinese,yangPennyYourThoughts2015, keller2017manipulate, ratkiewicz2011detecting}: an organizing entity seeks to amplify a targeted narrative by directing a group of deputies to endorse it on social media. In a related but much smaller case in the US, the Michael Bloomberg presidential campaign hired social media workers to post campaign messages~\cite{latimes_2020}. Twitter suspended their accounts, explaining that they had violated the Twitter Rules against platform manipulation and spam, and stating that the company would stop all who behave in ``substantially the same manner''~\cite{latimes_2020}. 

Twitter's rules forbid ``coordinating with or compensating others to engage in artificial engagement or amplification.'' But they allow ``coordinating with others to express ideas, viewpoints, support, or opposition towards a cause''\footnote{\url{https://perma.cc/Y2SX-BF3P}}. With these vague definitions, Twitter attempts to draw a line between coordination for ``authentic engagement'' and ``artificial amplification''. The policy hinges on a notion of a valid cause that justifies coordinated expression. In the campaigns we observed, at least the occasional participants can be considered genuine supporters of a political cause. Their cause could be right-wing, Hindu-nationalist, and even in contradiction with pluralistic and democratic principles. But there is little doubt that their support is genuine, which differentiates the campaigns from the astroturfing activities by the Bloomberg presidential campaign or the 50 Cent Army~\cite{yangPennyYourThoughts2015, king2017chinese}.

Yet, \textit{trend alert} campaigns are not grassroots movements, even if they draw on genuine citizen support. The campaigns lack the spontaneity and self-organizing capacities described by ``Social Movement Theory 2.0'' theorists~\cite{shirky2008here}. Instead, a political background organization gathers supporters around collective identities and asks them to volunteer their voices for a strategic action program. Is this authentic political participation? As Starbird et al.~\cite{starbird2019disinformation} argue, authentic and orchestrated expressions blur when citizen voices take on themes promoted by background actors.

Whether \textit{trend alert} campaigns are publicity or propaganda requires more than a judgment of authenticity. The campaigns may enable new forms of political expression and reinforce elite institutional politics while drowning more substantive citizen participation. A policy response will require a judgment of what constitutes legitimate political participation in India \cite{gillespie2010politics} and will depend on the perspective of the person making the assessment \cite{jack2017lexicon}. India's ruling political party seems to benefit from the campaigns and may have little interest in a critical evaluation of their legitimacy. As of August 2020, we have observed no significant platform action against users who participated in \textit{trend alert} campaigns. 

\subsection{Political media manipulations across platforms and borders}
The absence of platform action should not obscure that American technology companies affect who emerges as powerful in India's democracy. With its trend list, Twitter claims to provide a measure of the public mind similar to opinion polls \cite{gillespie2012can}. Various parties have tried to game the algorithm \cite{zhang2016twitter, nimmo2019measuring} that not only pushes narratives into public view but it also endorses their legitimacy. When Twitter turns a hashtag into a trend, it takes part in the contested process of political representation. Our data shows that 69 out of the 75 \textit{trend alert} campaigns successfully created a nationwide Twitter trend, often within minutes after launch. While we cannot measure the campaigns' effects on election outcomes, some led to further press coverage.

The activities described in this paper investigate media manipulations in the Indian political context. They also contribute more widely to an understanding of how political organizations worldwide use technologies to win elections. Tactical and structural innovations of novel campaigns often provide a proto-organizational example guiding waves of later innovations by political parties around the globe~\cite[p. 78]{karpf2012moveon}. In the US, the Patriot Journalist Network, a group of conservative activists, has been found to use a similar strategy of coordinating hashtag rallies using a dedicated campaign app to post pre-written tweets~\cite{mak_2017}. Multiple senators and members of Congress have extended their gratitude to the group for their social media support, but Twitter has decided to shut down the group's app based on a violation of its spam policy~\cite{mak_2017}.
 
Single-platform responses may fail when political campaigns combine old and new technologies, tactics, and organizational forms. A study of WhatsApp groups could not have shown how the campaigns were put into action, how large they were and whether they succeeded. A Twitter-specific analysis could not have ascertained the coordination behind the activities. Enforcement focused on a single platform will struggle to capture infringements by hybrid, cross-platform collectives. While the focus on a single platform is understandable from a pragmatic point of view, it may lend itself to overly simplistic understandings of information operations \cite{starbird2019disinformation}. We thus conclude this paper urging researchers and practitioners to pay attention to political information operations in the world's largest democracy and to support a debate towards practices that enable research and policy across multiple platforms. 

\subsection{Limitations and reflections}
To limit the scope of the data gathering and analysis, we focused this study on the WhatsApp group associated with the BJP, the most prominent Indian political party with the most expansive social media presence \cite{pal2019india}. We know that other Indian parties are using similar strategies, as we found variations of \textit{trend alerts} and \textit{tweet banks }in the WhatsApp groups of the Indian National Congress. Analyzing the other parties' strategies will contribute to a more comprehensive picture of media manipulations in India. 

Our analysis only considered one configuration of a social media manipulation strategy. Trend alerts were likely one of many parallel influence vectors in a massive election campaign. We missed coordination activities outside the WhatsApp groups we joined; we did not analyze the official NaMo app and had no insight into regional IT cells.  We also did not consider the \textit{image banks}~\cite{garimella2020images} that organizers distributed in the WhatsApp groups. 

To be transparent, we are concerned about democratic backsliding in India \cite{khaitan2020killing, goel_gettleman_khandelwal_2020, ganguly_9247}. However, we tried to separate our political views from the description of the campaigns. We believe it is essential to shedding light on emerging political practices,  so policy makers, platform operators, and a wider public can evaluate the campaigns' legitimacy.

\section{Conclusion}
We have documented a cross-platform media manipulation campaign with a novel configuration of technologies and organizational forms used in the 2019 Indian General Election. Organizers relied on an extensive network of WhatsApp groups aggregating loosely affiliated political supporters to coordinate controlled mass-posting on Twitter. Our findings suggest that a hybrid cross-platform organization produced hundreds of nationwide Twitter trends. Centrally controlled but voluntary in participation, this novel configuration of a  campaign complicates the debates over the legitimate use of digital tools for political participation. Trend alerts may have provided a blueprint for participatory media manipulation by a party with popular support.
\section{Appendix}

\subsection{List of BJP-affiliated Twitter accounts}
\n{To facilitate future work, we provide Twitter handles of BJP-affiliated accounts we followed. The list built on prior work by \citet{tulasi2019catching} and was manually verified to represent official party accounts or accounts of elected BJP representatives. Accounts that posted tweets with content found in \textit{tweet banks} are underlined with brackets indicating the number of suspicious tweets.  For a more exhaustive list of Indian politicians' Twitter accounts, refer to \citet{panda2020nivaduck}.}

\sloppy
\n{
@abhilashbjpmp, 
@abvpvoice, 
@advakash, 
@ahir\_hansraj, 
@ajaytamtabjp, 
\underline{@ajrajbjp~(54)},
\underline{@alongimna~(2)}, 
\underline{@amal\_mahadik~(2)}, 
@amaragrawalbjp, 
@amitmalviya, 
@amitshah, 
\underline{@amitthakerbjp~(1)}, 
@ananthkumar\_bjp, 
@anil\_baluni, 
@aniljaindr, 
@anilshirolebjp, 
\underline{@anilsole1~(2)}, 
@anupamtr, 
@anupmajaisbjp, 
@arunsinghbjp, 
@ashishsainram, 
\underline{@ashokgoelbjp~(1)}, 
\underline{@ashoknetemp~(1)}, 
@ashwinibjp, 
@atnanapatilmp, 
@avadhutwaghbjp, 
@balabhegade, 
@balbirpunj, 
@bharatendrabjp, 
@bharatpandyabjp, 
\underline{@bhatkhalkara~(3)}, 
@bhikhubhaidbjp, 
@bhupendrasingho, 
\underline{@bjp4andhra~(53)}, 
@bjp4ann, 
\underline{@bjp4arunachal~(11)}, 
\underline{@bjp4assam~(11)}, 
@bjp4bengal, 
\underline{@bjp4bihar~(1)}, 
@bjp4cgstate, 
@bjp4chandigarh, 
@bjp4chapra, 
@bjp4damandiu, 
\underline{@bjp4delhi~(1)}, 
\underline{@bjp4dnnh~(5)}, 
@bjp4goa, 
@bjp4gujarat, 
\underline{@bjp4haryana~(165)}, 
\underline{@bjp4himachal~(1)}, 
@bjp4india, 
\underline{@bjp4jharkhand~(157)}, 
\underline{@bjp4jnk~(1)}, 
@bjp4karnataka, 
\underline{@bjp4keralam~(3)}, 
@bjp4lakshadweep, 
\underline{@bjp4maharashtra~(1)}, 
\underline{@bjp4manipur~(2)}, 
\underline{@bjp4meghalaya~(1)}, 
\underline{@bjp4mizoram~(1)}, 
@bjp4mp, 
\underline{@bjp4nagaland~(2)}, 
@bjp4odisha, 
@bjp4puducherry, 
@bjp4punjab, 
@bjp4rajasthan, 
\underline{@bjp4sikkim~(1)}, 
@bjp4tamilnadu, 
@bjp4telangana, 
\underline{@bjp4tripura~(1)}, 
@bjp4uk, 
@bjp4up, 
@bjpbiplab, 
@bjpcpsingh, 
\underline{@bjpdilippatel~(4)}, 
@bjpdrmahendra, 
@bjplive, 
@bjpprakashmehta, 
@bjpramswaroop, 
\underline{@bjpsamvad~(1)}, 
@bjpsanjayjoshi, 
@bjpshivpshukla, 
@bjptarunchugh, 
\underline{@bjpvinodsonkar~(1)}, 
\underline{@bjym~(1)}, 
@blsanthosh, 
@bsybjp, 
\underline{@bunty\_bhangdiya~(1)}, 
@byadavbjp, 
@chandrabosebjp, 
@chaudhryshankar, 
@chdadapatil, 
\underline{@cpjoshibjp~(2)}, 
@crpaatil, 
@daddanmishra, 
@dangarbharat, 
@darshanajardosh, 
@dattatreya, 
@dev\_fadnavis, 
@devusinh, 
\underline{@dilipgandhimp~(1)}, 
@dilipghoshbjp, 
@dpradhanbjp, 
@dranilbondemla, 
\underline{@drashokuike\_mla~(1)}, 
@drashwathcn, 
@drdineshbjp, 
@drheena\_gavit, 
@drkathiria, 
@drlaxmanbjp, 
@drmnpandeymp, 
@drrganandbjp, 
\underline{@drrutvij~(1)}, 
@drtamilisaibjp, 
@dvsbjp, 
@eknathkhadsebjp, 
@fskulaste, 
\underline{@ganeshjoshibjp~(1)}, 
@ganeshsingh\_in, 
@gauravbh, 
@gautamgambhir, 
@girirajsinghbjp, 
\underline{@girishvyasbjp~(2)}, 
@gopalkagarwal, 
\underline{@gssjodhpur~(11)}, 
@gupta\_vijender, 
@guptaravinder71, 
@gvlnrao, 
\underline{@haribabubjp~(48)}, 
\underline{@harishd\_bjp~(2)}, 
@harishkhuranna, 
@hrajabjp, 
@ikjadejabjp, 
@imbhupendrasinh, 
\underline{@ipankajshukla~(9)}, 
@ishwarparmarmla, 
@jaggesh2, 
@janardan\_bjp, 
@jaunapuriass, 
@jayanta\_malla, 
@jayantsinha, 
@jhaprabhatbjp, 
@jigar\_inamdar, 
\underline{@jitu\_vaghani~(1)}, 
@jpnadda, 
@kailashonline, 
@kalrajmishra, 
\underline{@kamalsharmabjp~(1)}, 
\underline{@kapilpatilmp~(1)}, 
@keshavupadhye, 
@khadseraksha, 
@khushsundar, 
@kiritsomaiya, 
@kirronkherbjp, 
@kishanreddybjp, 
@kpgbjp, 
\underline{@krishnakhopde~(1)}, 
\underline{@kuljeetschahal~(7)}, 
@lallusinghbjp, 
@lavekarbharati, 
@lkbajpaibjp, 
@m\_lekhi, 
@madhavbhandari\_, 
@madhukeshwar, 
@madhurimisal, 
\underline{@maheishgirri~(1)}, 
@maheshpoddarjhr, 
@manekagandhibjp, 
@mangalpandeybjp, 
@manojsinhabjp, 
@manojtiwarimp, 
@mansukhmandviya, 
@medha\_kulkarni, 
@mepratap, 
@mlajagdishmulik, 
\underline{@mohitbharatiya\_~(1)}, 
@mp\_meerut, 
@mpharishchandra, 
@mplodha, 
@mprakeshsingh, 
@mpvadodara, 
@mrsgandhi, 
@msufisaint, 
@mukeshpatelmla, 
@nalinskohli, 
@naqvimukhtar, 
@narendramodi, 
@narinderbragta, 
@nirmalsinghbjp, 
@nitin\_gadkari, 
@nityanandraibjp, 
@nkishoreyadav, 
@npanchariyabjp, 
@nupursharmabjp, 
@ommathur\_bjp, 
\underline{@opsharmabjp~(1)}, 
\underline{@p\_sahibsingh~(2)}, 
@pankajbhaidesai, 
@pankajsinghbjp, 
@parag\_alavani, 
@parinayfuke, 
@pavanranarss, 
@pcmohanmp, 
@pemakhandubjp, 
@pmuralidharrao, 
\underline{@ponnaarrbjp~(8)}, 
\underline{@pradipsinhbjp~(1)}, 
@pratyushkanth, 
@pvssarma, 
@radhamohanbjp, 
@rahulkaswanmp, 
@rahulkotharibjp, 
@rahulsinhabjp, 
@rajeev\_mp, 
@rajeevbindal, 
@rajengohainbjp, 
@rajeshchudasma, 
@rajivprataprudy, 
@rajnathsingh, 
@rajnipatel\_mla, 
\underline{@rakeshshahmla~(4)}, 
@ramadevibjp, 
@rameshwar4111, 
@ramkripalmp, 
@ramlal, 
@rammadhavbjp, 
@ramvicharnetam, 
\underline{@raosahebdanve~(2)}, 
@rbsharmabjp, 
@rksinhabjp, 
\underline{@rohit\_chahal~(27)}, 
@rshuklabjp, 
@rspandeybjp, 
@sambitswaraj, 
\underline{@sanghaviharsh~(3)}, 
@sanjaydhotremp, 
\underline{@sanjaytandonbjp~(4)}, 
@santoshranjan\_, 
@sarojpandeybjp, 
@satpalsattibjp, 
@satyapaljain, 
@sbhagatbjp, 
@sbhttachrya, 
@shabdsharanbjp, 
@shahnawazbjp, 
@shainanc, 
\underline{@sharmakhemchand~(3)}, 
@shatrugansinha, 
@shelarashish, 
@shivprakashbjp, 
\underline{@shobhabjp~(3)}, 
@shweta\_shalini, 
@shyamsjaju, 
@siddharthanbjp, 
@sriramulubjp, 
@ssahluwaliamp, 
@subhashbrala, 
@sudhanshubjp, 
@sudhirparwemla, 
\underline{@sujit\_thakurmlc~(2)}, 
\underline{@sunil\_deodhar~(10)}, 
@sunilambekarm, 
\underline{@sunilbansalbjp~(1)}, 
@sunilpdeshmukh, 
\underline{@sunilsingh\_bjp~(7)}, 
@sureshbhattbjp, 
@sureshhalwankar, 
@sureshkpujari, 
\underline{@sureshranabjp~(1)}, 
@suryahsg, 
@sushmaswaraj, 
@swamy39, 
\underline{@swatantrabjp~(2)}, 
@tajinderbagga, 
@tapirgao, 
@tarunvijay, 
@tejasvi\_surya, 
@theashoksinghal, 
@tigerrajasingh, 
@tinkuroybjp, 
@tsrawatbjp, 
@upadhyaysbjp, 
@urspmr, 
@v\_shrivsatish, 
@vanathibjp, 
@varshabendoshi, 
@vasundharabjp, 
@vijayarahatkar, 
@vijaygoelbjp, 
@vijaypalbjp, 
@vijaypathakbjp, 
\underline{@vijayrupanibjp~(2)}, 
@vijeshlunawat, 
@vinay1011, 
@vinitgoenka, 
@vinodchavdabjp, 
@vipulgoelbjp, 
@virenderbjp, 
\underline{@vmbjp~(5)}, 
@zankhanabenbjp.}

\bibliography{references} 
\bibliographystyle{ACM-Reference-Format}

\end{document}